\documentclass[journal]{IEEEtran}
\pdfoutput=1

\usepackage{ifpdf}
\usepackage{subfigure}
\usepackage{graphicx}
\DeclareGraphicsExtensions{.pdf,.jpg,.png}
\usepackage[cmex10]{amsmath}
\usepackage{amssymb}
\usepackage{array}
\usepackage{mdwmath}
\usepackage{mdwtab}
\usepackage{pslatex}
\usepackage{xcolor}
\usepackage{textcomp}
\usepackage{placeins}
\usepackage{eqparbox}
\usepackage{url}
\usepackage{stfloats}
\usepackage{multicol}
\usepackage{subfigure}
\usepackage{float}
\usepackage{algorithmicx}
\usepackage{balance}
\usepackage{algpseudocode}
\usepackage{amsthm}
\usepackage{comment}

\begin{document}

\title{Last Meter Indoor Terahertz Wireless Access: Performance Insights and Implementation Roadmap}

\author{
\IEEEauthorblockN{Vitaly Petrov\IEEEauthorrefmark{2}, Joonas Kokkoniemi, Dmitri Moltchanov, Janne Lehtom{\"a}ki, Yevgeni Koucheryavy, Markku Juntti\vspace{-8mm}}
\thanks{V. Petrov, D. Moltchanov, and Y. Koucheryavy are with the Laboratory of Electronics and Communications Engineering, Tampere University of Technology, Tampere, Finland.}
\thanks{J. Kokkoniemi, J. Lehtom{\"a}ki, and M. Juntti are with the Centre for Wireless Communications, University of Oulu, Oulu, Finland.}
\thanks{The work of V. Petrov, D. Moltchanov, and Y. Koucheryavy was supported by Academy of Finland. The work of J. Kokkoniemi, J. Lehtom{\"a}ki, and M. Juntti was supported by Horizon 2020, European Union's Framework Programme for Research and Innovation, under grant agreement No. 761794 (TERRANOVA project). V. Petrov also acknowledges the support by Nokia Foundation and HPY Research Foundation funded by Elisa.}
\thanks{Manuscript submitted March 2016; revised January 2017, October 2017, and January 2018; accepted February 2018.}
\thanks{\IEEEauthorrefmark{2}V. Petrov is the contact author: vitaly.petrov@tut.fi}
}

\maketitle

\begin{abstract}
The terahertz (THz) band, 0.1--10\,THz, has sufficient resources not only to satisfy the 5G requirements of 10\,Gbit/s peak data rate but to enable a number of tempting rate-greedy applications. However, the THz band brings novel challenges, never addressed at lower frequencies. Among others, the scattering of THz waves from any object, including walls and furniture, and ultra-wideband highly-directional links lead to fundamentally new propagation and interference structures. In this article, we review the recent progress in THz propagation modeling, antenna and testbed designs, and propose a step-by-step roadmap for wireless THz Ethernet extension for indoor environments. As a side effect, the described concept provides a second life to the currently underutilized Ethernet infrastructure by using it as a universally available backbone. By applying real THz band propagation, reflection, and scattering measurements as well as ray-tracing simulations of a typical office, we analyze two representative scenarios at 300\,GHz and 1.25\,THz frequencies illustrating that extremely high rates can be achieved with realistic system parameters at room scales.
\end{abstract}
\vspace{-0.1cm}
\begin{IEEEkeywords}
Terahertz band communications, beyond-5G networks, massive data offloading, last-meter connectivity, ray-based modeling, THz band propagation measurements.
\end{IEEEkeywords}

\vspace{-0.1cm}
\section{Introduction}


The ever-increasing wireless data demands place extreme requirements on the future communications systems. In addition to physical layer improvements including advanced coding, massive multiple-input and multiple-output (massive MIMO), and cognitive radio systems, researchers currently investigate a number of architectural solutions. Since most of the traffic is generated indoors, the future systems are expected to rely on a significant number of indoor small cells to \emph{massively offload heavy traffic} from cellular networks. To enable this, the millimeter wave (mmWave) frequencies such as $28$\,GHz, $60$\,GHz have been recently investigated. However, the use of mmWaves still leads to certain limitations as the shared throughput indoors will only approach a few Gbit/s.


Several wireless communication actors are already investigating the use of even higher frequencies available in the terahertz (THz), $0.1$--$10$\,THz, frequency band, e.g. $120$\,GHz and $300$\,GHz and even the entire THz band~\cite{Jornet2011}. With extreme antenna gains, a number of testbeds operating in the lower THz band have already demonstrated data rates on the order of tens of Gbit/s over hundreds of meters as in~\cite{kallfas}. In indoor environments with mobility of users, realistically, the THz band is the most suitable for short link distances such as few meters. The achievable capacity at such distances could be extremely high, approaching hundreds of Gbit/s~\cite{Jornet2011,boronin}.


The potential of THz band has been recently recognized by IEEE, which formed an IEEE 802.15.3d Task Group identifying THz band communications as a feasible wireless technology for extremely high access rates of up to $100$\,Gbit/s. If properly integrated into the existing infrastructure, a wireless communications solution operating in the THz band could become a technology enabler for massive data offloading, satisfying $10$\,Gbit/s peak data rate requirement of 5G systems, as well being the first competitor to wired Internet access potentially providing comparable data rates and latency.


Together with the exceptional promises, THz communications bring novel unique challenges requiring to rethink the classic communications mechanisms. The ultra-wideband extremely directional nature of the communications links leads to the fundamentally new propagation and interference structures in a system with multiple reflected, diffracted and scattered beams causing complex received signal waveform. Medium access protocols have to operate with narrowly focused beams, fast handover procedures have to include the time required for localization and tracking functionalities. This combination of challenges is unique to the THz systems and has never been observed at lower frequencies. Finally, the question of enabling an ultra-high speed wireless access is related to finding cost-effective solutions for connecting it to the Internet. This is foreseen as a hidden bottleneck of wireless systems that may strike in the near future.


In this article, we present the roadmap towards an indoor THz communications technology well integrated into the existing infrastructure. Specifying the steps needed to achieve the goals and related challenges allows to unify the efforts of the community to build a solid way towards networking in the THz band. We also perform an extensive joint measurement-simulation campaign to report the capacity and signal-to-noise ratio (SNR) in typical indoor scenarios for the existing level of electronics and transceivers. 

The attractive values of estimated capacity and data rates confirm the claim that the successful implementation of THz wireless access will provide the bearer technology not only for conventional Internet access and massive data offloading but to a number of emerging applications requiring exceptionally high data rates in indoor scenarios, such as augmented reality, mobile-edge computing, and holographic communications.


The article is organized as follows. In the next section, we describe the state-of-the-art in THz communications highlighting the lessons learned over the last two decades. Later on, we introduce the vision of the last-meter THz access and assess its potential for realistic indoor scenarios is performed next. The roadmap towards fully integrated THz systems is then outlined. Conclusions are drawn in the last section.

\section{THz Communications: The Lessons Learned}\label{sec:lessons}

\subsection{THz Emitters}

One of the reasons for slow take-off of THz communications technologies has been the unique technical challenge, known as the \emph{THz gap}~\cite{song201450}. The THz frequency range is too high for the regular oscillators and too low for the optical photon emitters to let any of them be used as a THz signal generator. Thereby, THz waves are often generated by either a combination of lower frequency oscillator and frequency multiplier or an optical signal source (e.g., laser) and frequency divider. Both approaches have the overall efficiency several orders of magnitude lower than that of direct signal generator.

The applicability of both Si and SiGe technologies is limited to $\sim 200$--$300$\,GHz. This intrinsic device speed limitation is set by the transit time of the carriers through the material. Materials beyond the conventional Si and SiGe for THz applications include GaN, InP, and metamorphic technologies. These materials exhibit high electron mobility and large material breakdown voltage. InP-based and GaN HEMTs have been reported with a cut-off frequency greater than $600$\,GHz and maximum oscillation frequency higher than $1.2$\,THz. Resonant tunneling diodes (RTDs) are also actively studied for building THz semiconductor oscillators.

Current efforts to increase the output power at THz include the iBROW project targeting transmits more than $1$mW with RTD on a III-V with Si platform. Optoelectronic RTD could enable mmWave or THz femtocells connected to high-speed optical networks. Portable devices are pursued with all-electronic RTD\footnote{H2020 project ``Innovative ultra-BROadband ubiquitous Wireless communications through terahertz transceivers (iBROW)'': http://ibrow-project.eu/ [Accessed 10-02-2017]}. The DARPA THzE program target is $10$\,mW at $1.03$\,THz~\cite{Albrecht2011}. Finally, nanoplasmonic technologies, manipulating electromagnetic radiation at the scales smaller than the wavelength of light, and, thus, overcoming the diffraction limit, have recently attracted special attention~\cite{Jornet2013}.

\vspace{-0.2cm}
\subsection{THz Wave Propagation}

The abovementioned efforts promise to soon close the THz gap allowing for compact yet powerful THz transceivers. However, even if the THz signal can be generated and received, a challenge that still questions the applicability of the THz band for communications is the attenuation of the THz signal with distance. The frequency dependency of the free space path loss comes from the frequency dependent antenna aperture of the receiver. The effective antenna aperture is \emph{million times} lower at $1$\,THz than at $1$\,GHz, resulting in $60$\,dB higher attenuation. The transmit power at THz will likely also be lower. As a result, for reliable connectivity over a few meters, high gain antennas on both Tx and Rx are needed. This is feasible as high gain antennas are easier to construct at higher frequencies.

Molecules absorb electromagnetic energy at their resonance frequencies. Notable already for mmWaves, molecular absorbtion becomes much more harmful in the THz band. Following the Beer-Lambert law, the absorption loss is \textit{exponential} in the distance. To overcome this issue, the use of the sub-bands less affected by molecular absorption, so-called ``transparency windows" (e.g.  $0.1$--$0.54$\,THz), have been recently proposed~\cite{boronin}.

\begin{figure}[!t]
	\centering
	\includegraphics[width=1.0\columnwidth]{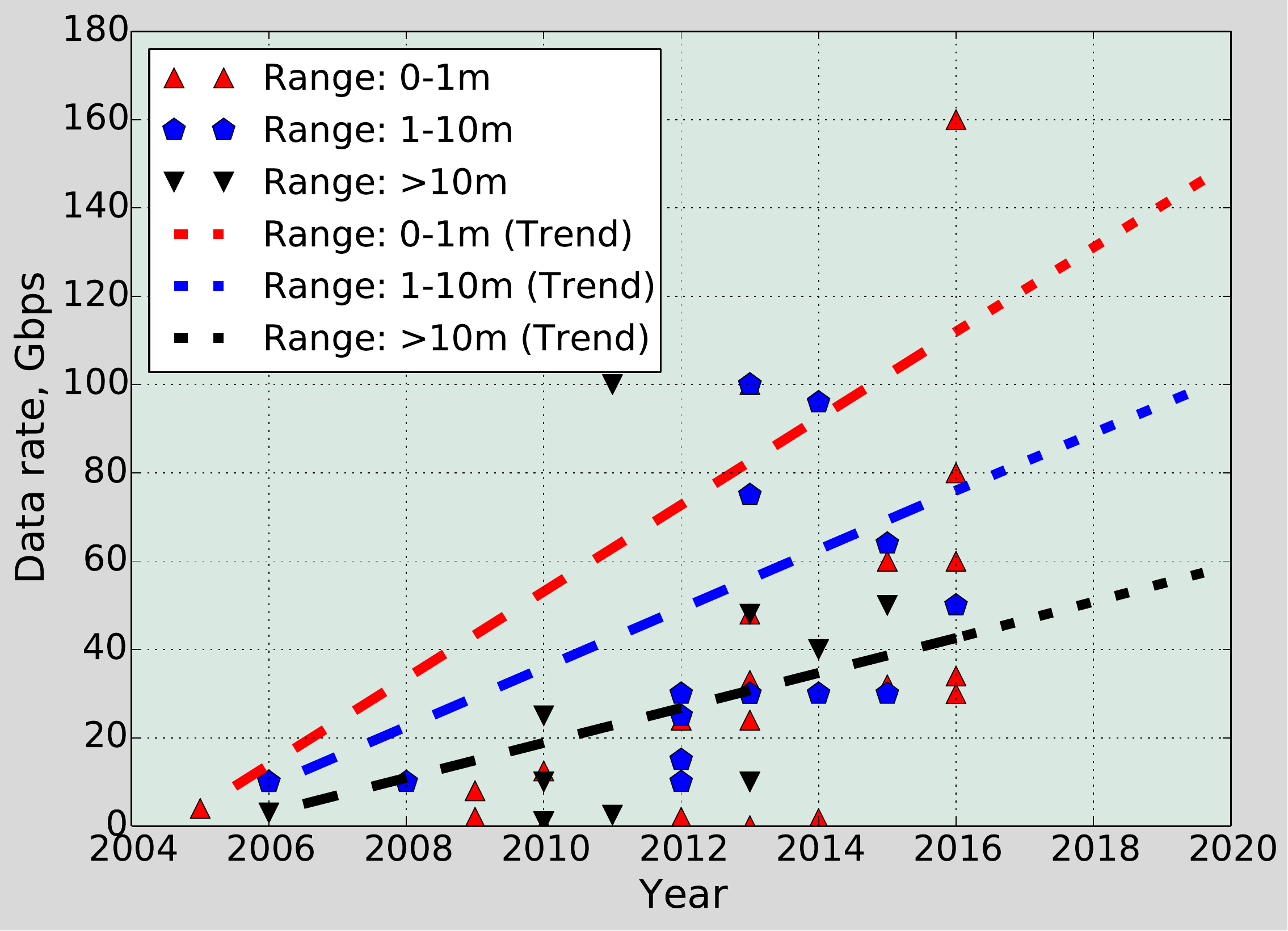}
	\caption{Advances in THz communications prototypes.}
	\label{fig:proto}
\end{figure}

\begin{table*}[!t]
\vspace{-0.2cm}
\renewcommand{\arraystretch}{1}
\caption[caption]{The prospective applications of THz wireless communications.\\Summary of IEEE 802.15.3\lowercase{d} vision (based on \cite{ieee_app_req}, \cite{ieee_tech_req}, and individual contributions)} \label{table:compare} \centering
{\footnotesize
\vspace{-0.2cm}
\begin{tabular}{|l|l|l|l|r|r|r|} \hline
Application & Topology & Antennas  directivity & Beam steering & Range, m  & Data rates, Gbit/s & Target BER\\
\hline \hline
Wireless backhaul & Point-to-point & Extremely directive ($>$50dBi each side)\qquad{}\qquad{}& Not required & $500$  & $10$--$100$ & $10^{-12}$\\ \hline
Wireless fronthaul & Ad hoc & Highly directive (Base station: $>$20dBi) & Mandatory & $200$ & $10$--$100$ & $10^{-12}$\\ \hline
Data centers & Point-to-point & Highly directive ($>$20dBi each side) & Not required & $100$ & $1/10/40/100$ & $10^{-12}$ \\ \hline
Kiosk downloading & Point-to-point & Directive ($>$10dBi kiosk side)& Optional & $0.1$ & $1$--$100$ & $10^{-6}$ \\ \hline
Intra-chip networks & Ad hoc & Deployment specific & Optional & $0.03$ & $1$--$100$ & $10^{-12}$\\ \hline
\end{tabular}}
\vspace{-0.3cm}
\end{table*}

The absorption of energy by the environment results in the so-called \textit{molecular noise} as the absorbed energy is released back to the environment. This noise has been studied over the last few years as its presence may complicate link and system level analyses~\cite{petrovTWC}. There is yet no definite conclusion about its effect as this theoretically predicted phenomenon has never been observed in real experiments \cite{molnoise}. However, the majority of the models predict its level to be proportional to the received power level, and, thus, rapidly decreasing with distance.

\subsection{THz Communication Prototyping}

Several testbeds have already been demonstrated, especially, in the lower end of the THz band. In~\cite{song201450}, quadrature phase shift keying (QPSK) modulator and demodulator have been demonstrated at $0.3$\,THz. A miniature $15$\,dBi gain antenna with maximum dimension of 5mm with easy to connect to package has also been reported~\cite{song201450}. By integrating antenna and other components, a compact THz module can be developed. Communication at $0.385$\,THz has been demonstrated in~\cite{R1} using photonics based emitter and electronic receiver using oscilloscope, $32$\,Gbit/s data rate at link distances of $0.4$\,m was achieved with QPSK. Recently, a $0.3$\,THz prototype receiver small enough to fit in a mobile phone was presented by Fujitsu~\cite{R6}. Communication distance is currently limited to about $1$\,meter only and data rates are in the order of $20$\,Gbit/s. At the same time, this data rate is an order of magnitude larger than what could be achieved with lower bands.

\textcolor{black}{Broadening the summary, Fig.~\ref{fig:proto} illustrates the evolution of the THz communications test benches. Based on the communication range, we group them into three categories and supplement each group with the ``averaged'' line, highlighting the trend. The data are partially reproduced from~\cite{janne_new_link} and supplemented with the recently published materials. Analyzing these data, we stress the following three main points. First, the achieved data rate is constantly growing for all demonstrated categories. Then, the growth rate is, expectedly, inverse proportional to the communication range. That is, the shorter communication range is, the faster the achieved data rate is. Finally, as one may observe, the growth of the middle line (``Range: $1$--$10$\,m'' that we concentrate on in this paper) allows us to expect the $100$\,Gbit/s solutions to \textcolor{black}{appear by} $2020$.}

These practical testbeds and the progress with RTDs illustrate the progress in THz technology and the possibility to soon convert the testbeds into an operational networking solution.

\subsection{IEEE THz Standardization Activity}

The process of standardizing THz communications technology started back in 2008, when a Terahertz Interest Group (IGthz) has been established under the IEEE 802.15 family. During 2008--2013 this group led by TUBS, NICT, Intel, NTT, and AT\&T studied the fundamental limitations and capabilities of THz band. In 2014, it has been reformed to an IEEE 802.13.3d Task Group on ``100G wireless'' (TG100G) aiming to develop a PHY-MAC layer solution for wireless communications in the lower THz band, $252$--$325$\,GHz,~\cite{ieee_tech_req}.

The IEEE vision on prospective applications for the THz communications and associated technical requirements are summarized in Table~\ref{table:compare}. The scenarios range from wired links replacement in future electronics (e.g., intra-chip communications) and data centers to backhaul and fronthaul links for beyond 5G mobile systems. The range of applicable communications distances is from $3$\,cm to $500$\,m. Surprisingly, the list of the applications, envisioned by IEEE 802.15.3d group, has a gap in the range $0.1$--$100$\,m. However, based on the trends in THz technology development and constantly increasing demands for higher rates at the air interface, we believe that indoor wireless access via the THz band in the range $0.1$--$10$\,m has to be also addressed as one of the potential applications. In the following section, we describe the roadmap towards the last-meter terabit-per-second wireless access and highlight the associated research and engineering challenges.

\section{The Last Meter THz Wireless Access Vision}\label{sect:last}

\subsection{The THz Plug Concept}


Motivated by the recent progress in THz transceiver design, in this section, we introduce a high data rate ``last-meter'' indoor THz communications system reusing the existing Ethernet infrastructure for Internet connectivity and massive traffic offloading from wireless local area and cellular networks. \textcolor{black}{The presented concept is well integrated into the network infrastructure and represents a disruptive shift from the current access systems leading to rapid performance improvements.}

\begin{figure*}[t!]
	\centering
	\vspace{-5mm}
	\includegraphics[width=1\textwidth]{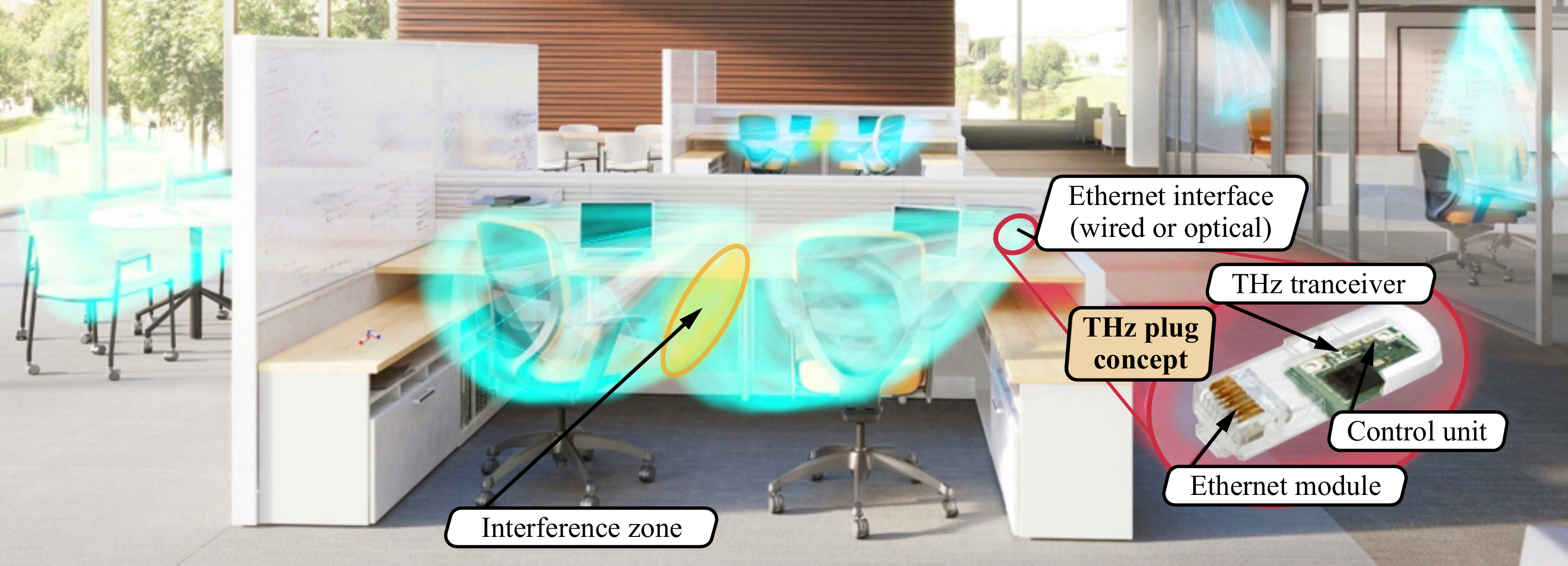}
	\caption{A vision of immersive THz indoor wireless access environment.}
	\label{fig:vision}
	\vspace{-5mm}
\end{figure*}

The envisioned \textit{THz plug}, see Fig.~\ref{fig:vision}, is a low-cost THz hotspot that is inserted into an Ethernet socket connected to standard Ethernet infrastructure and, if copper wire medium is used, powered via Power-over-Ethernet (PoE) technology. If a THz plug is placed next to the office desk, a user can reliably connect a laptop or tablet to a \emph{multi-gigabit-a-second} wireless link on any location on the table or few meters around.


The Ethernet infrastructure is already deployed in the offices and some other venues, leading to efficient and cost-effective data offloading to/from the high data rate THz interface. Currently, the Ethernet infrastructure is severely under-utilized due to the popularity of WLAN technologies. The envisioned system helps users to avoid convenience/performance trade-off by combining advantages of both wired and wireless networks and to fully benefit from the extraordinary rates offered by the THz frequency band. In conjunction with the modern Ethernet backbone, it enables truly broadband wireless indoor access.

Indoor locations comprise difficult propagation environment for the radio waves. The architectural solutions vary significantly from building to building and even from room to room. Potentially, a large number of furniture combined with walls and moving objects, such as doors and even humans, cause significant shadowing, thus, large spatial variations in the signal quality. The conventional indoor communication systems, such as WLANs, cope with these problems with sufficient penetration properties provided by the microwave band. Moving to the THz band, the objects become opaque and even small structures, e.g., a mug on a table may prevent the line-of-sight (LoS) communications, the lifeline of the THz communications. To assess the potential of the proposed concept we need to understand the principal benefits and limitation of THz propagation in indoor environment.


Owing to the complexity of the indoor propagation environment, we apply a two-stage measurement-simulation campaign to assess the performance of the proposed system:
\renewcommand{\theenumi}{\Alph{enumi}}
\begin{enumerate}
\item \emph{Field Measurement Campaign.} In the first stage, we measure THz waves propagation ($0.3$--$3$\,THz) in typical indoor environment focusing on waves reflection and scattering from such materials as concrete, plastic, hardwood, etc. To the best of authors' knowledge, the exact values for above $1$\,THz are reported for the first time.
\item \emph{Ray-Tracing Assessment.} In the second stage, we utilize the measured data to parameterize our ray-tracing simulation tool. We then simulate the typical office scenarios characterizing the performance of THz wireless access using capacity and SNR as metrics of interest.
\end{enumerate}

\vspace{-0.1cm}
\subsection{Field Measurement Campaign}

\begin{figure*}[!t]
\centering
	\vspace{-0.5cm}
	\includegraphics[width=1\textwidth]{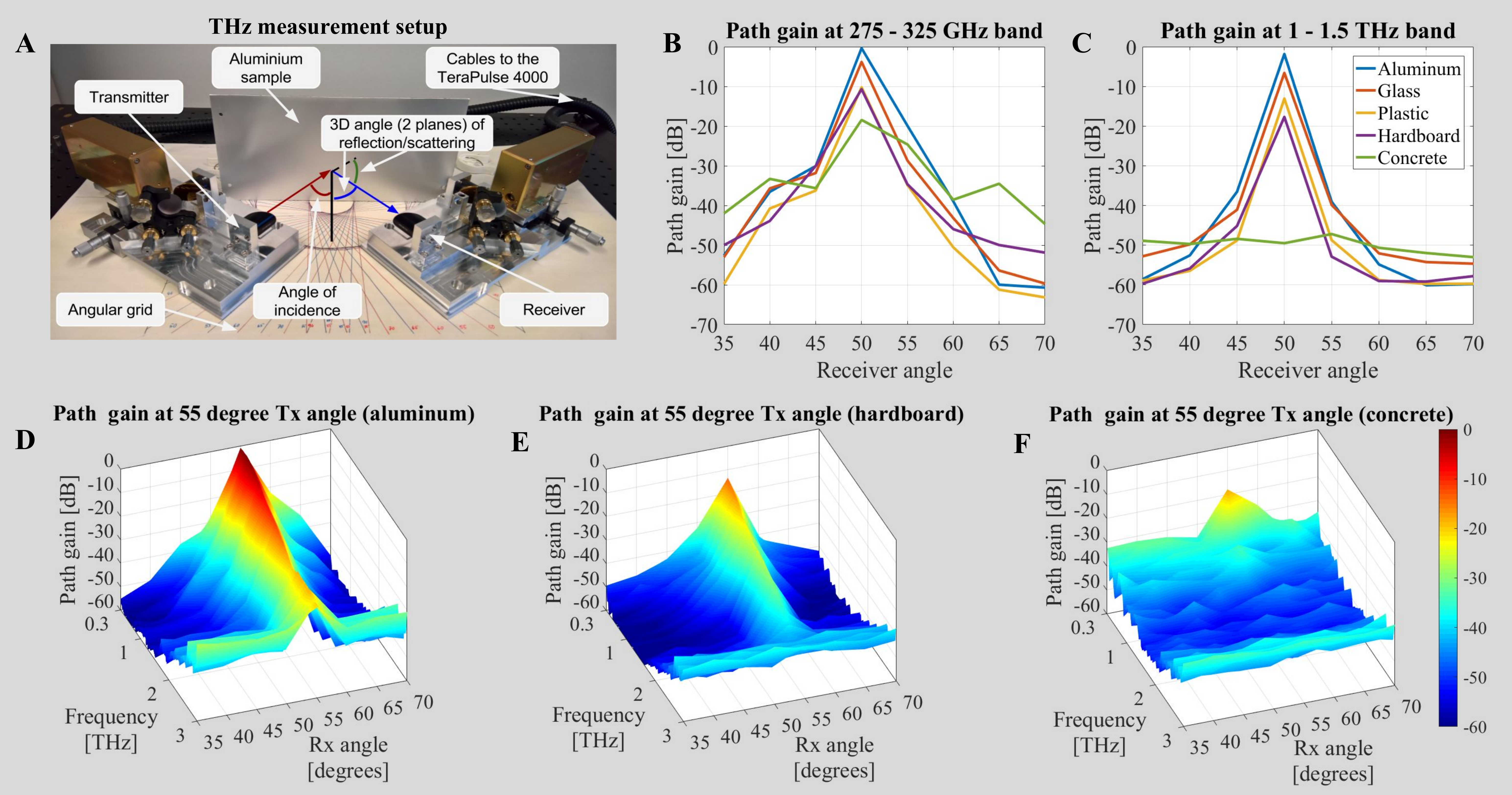}
    \caption{Measured scattering properties of aluminium, glass, plastic, hardboard, and concrete in $0.1$--$3$\,THz band.}
    	\vspace{-0.4cm}
    \label{fig:scattering}
\end{figure*}

We utilized TeraView Mini Pulse\footnote{TeraView, ``TeraPulse 4000 -- THz Pulsed Imaging and Spectroscopy'', http://www.teraview.com/products/TeraPulse\%204000/index.html [Accessed 10-02-2017]}, capable on THz band transmission/reception from $60$\,GHz to $4$\,THz, with time resolution of $8.3$\,fs and frequency resolution $5.9$\,GHz. Focus was on scattering properties of typical materials in office rooms including glass, plastic, hardwood, concrete and aluminium. \textcolor{black}{To obtain these data, two test benches have been developed. The main one is illustrated in} Fig.~\ref{fig:scattering}A \textcolor{black}{and is used to measure the amount of the received energy reflected/scattered from a given sample subject to an angle of incidence and a 3D angle of reflection/scattering. The second test bench presents a LoS transmission of the same signal over the equal distance but without any obstacle on the way. Finally, the measured values of received energy coming from the first test bench are normalized with the data obtained with the second one to eliminate the propagation effects. As the result, the virgin data on the materials reflection and scattering properties are obtained and reported in Fig.}~\ref{fig:scattering}.


\textcolor{black}{As one may observe from Fig.}~\ref{fig:scattering}, the smoother the material is, the higher is the peak response around the reflection path. The surface roughness decreases the reflected path energy by distributing the energy on the diffuse scattering field. The measurement results in Fig.~\ref{fig:scattering} show that the reflected path has the most energy, but also confirm that rough surfaces such as concrete have quite a flat response over the Rx angles. It can be seen that aluminium is the best reflector among the considered materials. However, glass, plastic, and hardboard are not much worse. One of the reasons is that THz signals do not penetrate aluminium, whereas  plastic for example does allow a part of the THz signal to go through. On the other hand, concrete has significantly different properties than the other materials. We can see from Fig.~\ref{fig:scattering}\,F that only at relatively low frequencies does concrete have a significant reflected component. This represents a challenge on THz modeling, both strongly reflecting and strongly diffusing materials must be considered. \textit{We would like stress that even at extremely high frequencies such as 3THz there is still a strong reflected component out of typical office materials.} This implies that nLoS communications through the first reflection might be feasible even at the THz frequencies.

\subsection{Ray-Tracing Assessment}

To assess performance of the prospective THz plugs in indoor environment, we designed ray-tracing simulation framework, based on the surface tessellation to miniature segments. The size of each segment is comparable to the wavelength, thus, the segments can be considered as point transceivers, receiving some energy from Tx and reflecting a part of it to Rx or to another point transceiver on a different surface.

\begin{figure*}[t!]
\centering
	\vspace{-0.5cm}
	\includegraphics[width=1\textwidth]{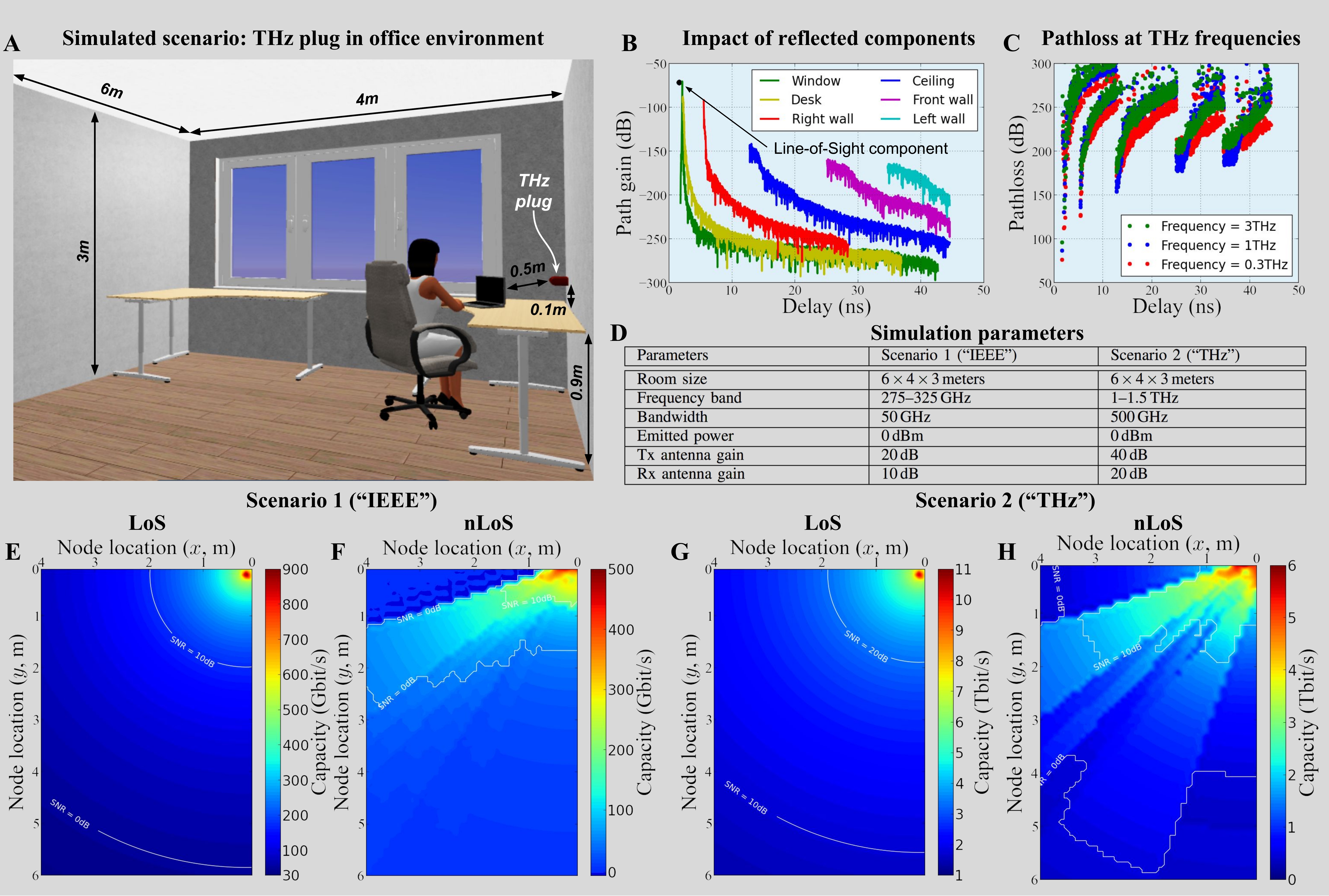}
    \caption{Specifics of THz room propagation for the two considered scenarios.}
        	\vspace{-0.4cm}
    \label{fig:thz_room}
\end{figure*}

We study a typical $6\times4\times3$ meters office room with walls, window, desks, ceiling and floor built from different materials as illustrated in Fig.~\ref{fig:thz_room}\,A. We consider two scenarios as shown in Fig.~\ref{fig:thz_room}\,D. The first one reflects the IEEE standardization efforts and takes into account the current state of THz electronics. With the second scenario, we take a look in the future, assessing performance of THz indoor communications by taking into account the potential progress in THz devices. Both LoS and nLoS communications are of interest. \textcolor{black}{The nLoS setup presents an ideal case, where the path having the lowest losses is always selected}.

We consider both laptop and mobile device connectivity cases. In the former case, the device is located at the wooden desk $10$\,cm below and $50$\,cm apart from the THz plug. In mobile connectivity case, a device might be located at any point within the room at a height of $1$\,m. The metrics of interest are capacity and SNR coverage of a room as well as power delay profile (PDP) of surfaces contributing to the received signal at a fixed point in a room.


PDP for involved surfaces in laptop connectivity case at $0.3$\,THz is shown in Fig.~\ref{fig:thz_room}\,B. As we observe, the LoS component is dominating and the reflections from the closest two surfaces (wooden desk and window) are $\sim20$\,dB weaker. The reflected rays from the concrete walls scatter a lot and are from $75$ to $120$\,dB lower than the LoS component, making their impact negligible. Notice that the level of scattering significantly affects the PDP structure from a selected surface.


The aggregated PDP is presented in Fig.~\ref{fig:thz_room}\,C. It is important that the same qualitative picture is observed for all the chosen frequencies, irrespective of whether they belong to transparency window ($0.3$\,THz) or not ($1$\,THz and $3$\,THz). As expected, the LoS is dominating the PDP. Then, there are clearly five peaks in the received signal (first reflections from the surfaces) each with its own tail of scattered waves, potentially enabling nLoS communication.


We show the average capacity for ``IEEE'' scenario between a THz plug and a mobile node located inside the room for LoS, Fig.~\ref{fig:thz_room}\,E, and nLoS, Fig.~\ref{fig:thz_room}\,F, cases. For nLoS we assume an obstacle close to the mobile node blocking the LoS, e.g. head, hand, etc. As one may observe, it is possible to ensure a reliable connectivity even in the absence of the LoS by capturing the power of the the best direct reflection: the minimal SNR is around $0$\,dB within $1$--$3$ meters from a THz plug (Fig.~\ref{fig:thz_room}\,F). This result assumes that the reflected component from the window is not blocked. The approach requires precise tracking of the node location and ability to dynamically adjust the transmit and receive antenna patterns on a THz plug, which is a critical research problem. At the same time we maintain a sufficiently high data rate, as the theoretical Shannon channel capacity is above $100$\,Gbit/s. Assuming pessimistic 10\% of modulation and MAC efficiency, this would lead to expected throughput of around $10$\,Gbit/s.


Finally, we present the results for the ``THz'' scenario assuming efficient directional antennas at both THz plug and mobile node. Observing Fig.~\ref{fig:thz_room}\,G and Fig.~\ref{fig:thz_room}\,H, the level of SNR is much higher for both LoS and nLoS even for ten times wider bandwidth and higher propagation loss. Although the concept of utilizing razor sharp beams might sound futuristic today, it illustrates the potential of the research in this field, as the theoretical link capacity achieves \emph{terabits-per-second}.

\section{Implementation Roadmap}

The proposed last-meter THz system allows for incremental development and can be specified in four phases with increasing complexity, usability and impact:
\renewcommand{\theenumi}{\Roman{enumi}}
\begin{itemize}
\item \emph{Phase I. Connectivity.} The Ethernet wireless extension system for point-to-multipoint connectivity with semi-stationary devices will be developed.
\item \emph{Phase II. Handover.} In-room nodes mobility and multiple THz plugs are included to the system. The emphasis is on enabling THz beamsteering at the air interface and associated mechanisms for handover support.
\item \emph{Phase III. Interference mitigation.} The support for multiple transmitters and receivers in the same environment is incorporated. Particularly, interference problems due to concurrent transmissions are to be addressed.
\item \emph{Phase IV. Integration.} THz wireless access is integrated as one of the radio access technologies into the Heterogeneous Network (HetNet) concept under the umbrella of beyond 5G systems.
\end{itemize}

\subsection{Phase I: Wireless Ethernet Extension}


In the first phase, directional antennas, beam learning/tracking, modulation, coding, and plug and play operation are addressed. Directional antennas (implemented e.g., with a phased array) are needed to increase the receiver's effective antenna aperture. The transmitter also needs to use directional antennas to compensate for the low transmit power. At millimeter waves, miniature phased arrays have already been implemented, but the achieved progress in THz is still limited and more work needs to be done. Going in this direction, the small wavelengths in the THz band enable packing tens of thousands of upcoming plasmonic nanoantenna elements in a small area~\cite{Jornet2013}. This is to be sufficient to support meters-distance links.

It may well be worth it to include several antenna arrays in \emph{THz plug} and mobile THz modules to reduce the chances for blocking and to support finding a good propagation path. With directional antennas, finding good beam pointing becomes critical. To support user mobility, fast and accurate beam tracking is required, \textcolor{black}{especially in nLoS conditions~\cite{new_one}}. Although mechanically steerable antennas for $0.3$\,THz have been demonstrated~\cite{firstDemo}, faster electronic beam steering is required for practical use.


For plug and play functionality the THz plug has to be seen by the network as a regular switch. Thus, the envisioned device is assumed to be IP-free with all the data forwarding performed on the link layer. A special question of interest is the possibility of preserving the Ethernet frame structure for seamless service extension to the air interface. The latter is highly desirable to avoid the continuous transcoding of both uplink and downlink traffic.

\subsection{Phase II: Point-to-Multipoint Functionality}


At the second stage, the user macro mobility has to be taken into account and supported with handover functionality. In particular, at this stage an SDN-like network controller, collecting data and managing all the THz plugs in a dedicated area, has to be introduced. The role of this device is to constantly monitor the user location and pre-select the new THz plug to catch the user, when it is close to leaving the current connection.

This type of feedback can be implemented by estimating user trajectories through beam configurations and received power level. Since the majority of the time the link is expected to be in LoS, and the THz plug is stationary with known location, the calculation is feasible. When a user is reaching the last available antenna configuration on its trajectory, the handover procedure has to be performed by directing the receiving beam from the next THz plug. Forwarding of unsent data from the current THz plug to the next one via the Ethernet interface has to be performed.

\subsection{Phase III: Multipoint-to-Multipoint Functionality}

Although the high directivity of beams may lead to noise-limited communications in the outdoor scenarios, the interference still plays a substantial role in the indoor environment. For instance, a massive interference might occur when two mobile Rx are in close proximity.

The experience accumulated standardizing IEEE 802.11ad highlights that completely new multiple access mechanisms will be required for systems with directional antennas. This is one of the fundamentally ``grey'' areas in the telecommunications research with no comprehensive solution proposed yet. On top of this, reflected/diffracted/scattered waves from one link may interfere with another, which is even more difficult to predict. Due to the presence of these effects in combination with nodes mobility, the total interference structure is completely new, in comparison to lower frequencies, with unpredicted reductions of received signal quality.

Therefore, an in-depth and accurate interference modeling for highly-directional THz antennas is required to understand and specify interference mitigation techniques for the multi-user THz wireless access. As candidates both conventional mitigation techniques such as power control and frequency division, as well as novel techniques, specific to the high-dense wireless networks and THz wireless access, need to be considered.

\subsection{Phase IV: Integration to HetNets}


Following the HetNet concept, currently envisioned as a collection of access networks with centralized control over the ''always-on'' macro base station interface, the proposed system can be integrated into the modern wireless world. Furthermore, the THz wireless access may become a network-controlled technology enabler for the recently introduced concept of Tactile Internet. Similarly to other access networks, the integration will require an in-building network control entity to interface with mobile operators infrastructure. The data plane will operate at high-speed last-meter connectivity whenever a user is in the proximity of a THz plug or being switched to a slower technology when a user leaves the coverage.

\section{Conclusions}


In this article, we introduced the concept of THz plug, a device operating in the terahertz frequency band, directly attached to an Ethernet socket. Combining gigabit Ethernet and the prospective THz wireless interfaces would enable massive wireless data transfers. A step-by-step roadmap towards the implementation of the proposed multi-gigabit last-meter indoor wireless access in the THz band is presented and major research challenges are highlighted.

We demonstrate the potential of the first step of the proposed roadmap for both IEEE 802.15.3d scenario ($50$\,GHz bandwidth with $300$\,GHz center frequency) and the true THz case with massive $500$\,GHz bandwidth in the $1$--$1.5$\,THz band. Despite the desirable capacity and data rate values, high level of THz signal attenuation and scattering is limiting the communication range and has to be mitigated in both LoS and nLoS cases by using sufficiently directive antennas with dynamic beamforming. Design of such antennas, associated transceivers, \textcolor{black}{as well as efficient beam tracking algorithms} is the key challenge to fully benefit from extraordinary capacity provided by the THz frequency band.

\balance
\bibliographystyle{ieeetr}
\bibliography{lastmeter}

{\small
\section*{Authors' Biographies}

\textbf{Vitaly Petrov} (vitaly.petrov@tut.fi) is a PhD candidate at the Laboratory of Electronics and Communications Engineering at Tampere University of Technology (TUT), Finland. He received the Specialist degree (2011) from SUAI University, St. Petersburg, Russia, as well as the M.Sc. degree (2014) from TUT. He was a Visiting Scholar with Georgia Institute of Technology, Atlanta, USA, in 2014. Vitaly (co-)authored more than 30 published research works on terahertz band communications, Internet-of-Things, nanonetworks, cryptology, and network security.

\textbf{Joonas Kokkoniemi} (joonas.kokkoniemi@oulu.fi) is a Postdoctoral Research Fellow with the Centre for Wireless Communications, University of Oulu. He received the B.Sc. (2011), M.Sc. (2012), and Dr.Sc. (2017) degrees from University of Oulu, Oulu, Finland. He was a Visiting Researcher with Tokyo University of Agriculture and Technology, Japan (2013) and a Visiting Researcher with State University of New York at Buffalo, USA (2017). Joonas's research interests include THz band and mmWave channel modeling and communication systems.

\textbf{Dmitri Moltchanov} (dmitri.moltchanov@tut.fi) is a Senior Research Scientist at the Laboratory of Electronics and Communications Engineering, Tampere University of Technology. He received his M.Sc. (2000) and Cand. Sc. (2002) degrees from Saint Petersburg State University of Telecommunications, as well as the Ph.D. (2006) degree from Tampere University of Technology. His research interests include performance evaluation and optimization issues in wired and wireless IP networks, Internet traffic dynamics, and traffic localization in P2P networks.

\textbf{Janne Lehtom{\"a}ki} (janne.lehtomaki@oulu.fi) is an Adjunct Professor with the Centre for Wireless Communications, University of Oulu. He received the M.Sc. (1999) and the Ph.D. (2005) in telecommunications from University of Oulu. His research interests are in terahertz wireless communication, channel modeling, IoT, and spectrum sharing. Janne co-authored the winner of the Best Paper Award at IEEE WCNC 2012. He is an Editorial Board Member of Elsevier Physical Communication.

\textbf{Yevgeni Koucheryavy} (evgeni.kucheryavy@tut.fi) is a Full Professor at the Laboratory of Electronics and Communications Engineering of Tampere University of Technology (TUT), Finland. He received his Ph.D. degree (2004) from TUT. His current research interests include various aspects in heterogeneous wireless communication networks and systems, the Internet of Things and its standardization, as well as nanocommunications. He is Associate Technical Editor of IEEE Communications Magazine and Editor of IEEE Communications Surveys and Tutorials.

\textbf{Markku Juntti} (markku.juntti@oulu.fi) received his Dr.Sc.\ (EE) degree from University of Oulu, Oulu, Finland in 1997. Dr.\ Juntti was with University of Oulu in 1992--98. In academic year 1994--95, he was a Visiting Scholar at Rice University, Houston, Texas. In 1999--2000, he was a Senior Specialist with Nokia Networks. Dr.\ Juntti has been a professor of communications engineering since 2000 at University of Oulu, Centre for Wireless Communications (CWC), where he leads the Communications Signal Processing (CSP) Research Group. He also serves as Head of CWC -- Radio Technologies (RT) Research Unit. His research interests include signal processing for wireless networks as well as communication and information theory. Dr.\ Juntti is also an Adjunct Professor at Department of Electrical and Computer Engineering, Rice University, Houston, Texas, USA. Dr.\ Juntti is an Editor of \textsc{IEEE Transactions on Communications}
}

\end{document}